\documentstyle[12pt,epsfig]{article}
\begin{document}
\def\mbn{\mbox{\boldmath$\nabla$}}
\def\intsum {\int\!\!\!\!\!\!\!{\ss \sum}}
\def\intsums {\int\!\!\!\!\!{\ss\Sigma}}
\def\rbh {\hat{\bf r}}
\def\Lb {{\bf L}}
\def\Xb {{\bf X}}
\def\Yb {{\bf Y}}
\def\Zb {{\bf Z}}
\def\rb {{\bf r}}
\def\kb {{\bf k }}
\def\qb {{\bf q}}
\def\kr {{\bf k}\cdot{\bf r}}
\def\pb {{\bf p}}
\def\etal{{\it et al., }}
\def\cf{{\it cf }}
\def\ie{{\em i.e., }}
\def\etc{ {\it etc}}
\def\dket#1{||#1 \rangle}
\def\dbra#1{\langle #1||}
\def\isim{\:\raisebox{-0.5ex}{$\stackrel{\textstyle.}{=}$}\:}
\def\lsim{\:\raisebox{-0.5ex}{$\stackrel{\textstyle<}{\sim}$}\:}
\def\gsim{\:\raisebox{-0.5ex}{$\stackrel{\textstyle>}{\sim}$}\:}
\def\a {{\alpha}}
\def\b {{\beta}}
\def\e {{\epsilon}}
\def\g {{\gamma}}
\def\r {{\rho}}
\def\s {{\sigma}}
\def\k {{\kappa}}
\def\l {{\lambda}}
\def\m {{\mu}}
\def\n {{\nu}}
\def\r {{\rho}}
\def\t {{\tau}}
\def\w {{\omega}}
\def\be{\begin{equation}}
\def\ee{\end{equation}}
\def\br{\begin{eqnarray}}
\def\er{\end{eqnarray}}
\def\brn{\begin{eqnarray*}}
\def\ern{\end{eqnarray*}}
\def\x{\times}
\def\go{\rightarrow  }
\def\rf#1{{(\ref{#1})}}
\def\nn{\nonumber }
\def\ket#1{|#1 \rangle}
\def\bra#1{\langle #1|}
\def\Ket#1{||#1 \rangle}
\def\Bra#1{\langle #1||}
\def\ov#1#2{\langle #1 | #2  \rangle }
\def\hf {{1\over 2}}
\def\hw {\hbar \omega}
\def\mbs{\mbox{\boldmath$\sigma$}}
\def\gsim{\:\raisebox{-0.5ex}{$\stackrel{\textstyle>}{\sim}$}\:}
\def\lsim{\:\raisebox{-0.5ex}{$\stackrel{\textstyle<}{\sim}$}\:}
\def\mbn{\mbox{\boldmath$\nabla$}}
\def\sss{\scriptscriptstyle}
\def\ss{\scriptstyle}
\def\endauthors{}
\def\authors#1\endauthors{#1}
\begin{titlepage}
\pagestyle{empty}
\baselineskip=21pt
\begin{center}
{\large{\bf Weak Magnetism in Two Neutrino Double Beta Decay}}
$^*$
\end{center}

\vskip .1in

\authors
\centerline{C.~Barbero${}^{ a\dagger}$,
F.~Krmpoti\'c$^{a\dagger}$, A. Mariano
$^{a\dagger}$
and D.~Tadi\'c${}^b$}
\vskip .15in
\centerline{\it ${}^a$ Departamento de F\'\i sica, Facultad de Ciencias
Exactas}
\centerline{\it
Universidad Nacional de La Plata, C. C. 67, 1900 La Plata, Argentina.}
\vskip .15in
\centerline{\it ${}^b$ Physics Department, University of Zagreb}
\centerline{\it Bijeni\v cka c. 32-P.O.B. 162, 10000 Zagreb, Croatia.}
\endauthors
\vskip 0.5in
\centerline{ {\bf Abstract} }
\baselineskip=18pt
We have extended the formalism for the two-neutrino double beta
decay by including the weak-magnetism term, as well as other
second-forbidden corrections.
The weak magnetism diminishes the calculated half-lives
in $\sim 10\%$,
independently of the nuclear structure.
Numerical computations were performed within the pn-QRPA, for $^{76}Ge$,
$^{82}Se$, $^{100} Mo$, $^{128}Te$ and $^{130}Te$ nuclei.
No one of the second-forbidden
corrections modifies significantly the spectrum shapes.
The total reduction in the calculated half lives varies from $6\%$ up to
$32\%$, and strongly depend on the nuclear interaction in the
particle-particle $S=1,T=0$ channel.
We conclude that the higher order effects in the weak Hamiltonian
would hardly be observed in the two-neutrino double beta experiments.

\vspace{0.5in}
\noindent
$^*$Work partially supported by
the CONICET from Argentina and the Universidad Nacional de La Plata.
\\ $^{\dagger}$Fellow of the CONICET
from Argentina.

\vspace{0.5in}
\end{titlepage}

\newpage
The sensitivity of the double-beta ($\beta\beta$) decay experiments is
steadily and constantly increasing. To become aware of this it suffices
to remember that, between the pioneer laboratory measurement on
$^{82}Se$ \cite{Ell87} and the most recent on $^{76}Ge$ \cite{Kla98},
the statistics has been improved by a factor of $\sim 5000$.

The quantity that is used to discern experimentally between the ordinary
Standard Model (SM) two-neutrino decays ($\beta\beta_{2\nu}$) and the
neutrinoless $\beta\beta$ events not included in the SM, both without
($\beta\beta_{0\nu}$) and with Majoron emissions ($\beta\beta_{\sss M}$),
is the electron energy spectrum $d\Gamma/d\epsilon$ of the decay rate
$\Gamma$. It is usually given as a function of the sum ($\epsilon$) of the
energies ($\epsilon_1$ and $\epsilon_2$) of the two emitted electrons.
The $\beta\beta_{2\nu}$ and  $\beta\beta_{\sss M}$ decays exhibit continuous
spectra in the interval $2\le\epsilon\le Q$, while the $\beta\beta_{0\nu}$
spectrum is just a line at the released energy $Q=E_{\sss I} -E_{\sss F}$.

In the evaluation of the $\beta\beta_{2\nu}$ decay rate,
the allowed (A) approximation is usually assumed to be valid, \ie
the Fermi (F) and Gamow-Teller (GT) operators are considered
at the same level of approximation as in single-allowed $\beta$ transitions.
Besides, as the F operator is strongly suppressed in the $\beta\beta_{2\nu}$
decay by the isospin symmetry, only the dominant contribution of the
axial-vector current is considered in practice. Then, the
$\beta\beta_{2\nu}$ matrix element, between the initial state
$\ket{0_{\sss I}}$ in the $(N,Z)$ nucleus to the final state
$\ket{0_{\sss F}}$ in the $(N-2,Z+2)$ nucleus (with energies $E_{\sss I}$
and $E_{\sss F}$ and spins and parities $J^{\pi}=0^+$) reads:
\br
{\cal M}_{2\n}^{\sss (0)}&=&
g_{\sss A}^2\sum_N\frac{\bra{0^+_{\sss F}}\mbs\ket{1^+_{\sss N}}\cdot
\bra{1^+_{\sss N}}\mbs\ket{0^+_{\sss I}}}{E_{1^+_N}-E_{0^+_I}+Q/2}.
\label{1}\er
Here the summation goes over the virtual states with spin and parity
$J^\pi=1^+$.

Except for the work of Williams and  Haxton \cite{Wil88}, all higher order
terms (HOT) in the weak Hamiltonian have been almost totally ignored in
the past, simply because they are expected to be small.
Yet, the question is not so simple. Firstly, from the comparison of the
recent experimental results for the $\beta\beta$
half lives in $^{76}Ge$ \cite{Kla98}: $T_{2\nu}\cong 1.77\cdot 10^{21}$ y,
$T_{\sss M}>1.67\cdot 10^{22}$ y and $T_{0\nu}>1.2\cdot 10^{25}$ y, it can be
stated that presently are being observed effects of the order of $10^{-4}$ at
$\epsilon \sim Q$ and of the order of $10^{-1}$ at $\epsilon \sim Q/2$.
Secondly, there are also several ongoing and planned experiments that are
supposed to allow for measurements of still smaller effects. Thus, the HOT
could, in principle, become relevant.

In recent years we have examined the effects of the first-forbidden operators
in the $\beta\beta_{2\nu}$ decay \cite{Bar95,Bar98a}, both the non-unique
(FFNU) and unique (FFU), which contribute via the virtual states
$J^\pi= 0^-,1^-$
and $J^\pi=2^-$, respectively. From these studies we have learned that:
a) the FFNU transitions might increase the $T_{2\nu}$ up to $\sim 30\%$,
yet they do not modify the A shape of the two-electron spectrum \cite{Bar95},
b) the FFU transitions do alter the $\beta\beta_{2\nu}$ spectrum shape but
only at the level of $10^{-6}$ and mainly at low two-electron energy, where
most backgrounds tend to dominate \cite{Bar98a}.
Therefore, the effects of the first-forbidden transitions could hardly screen
the detection of exotic $\beta\beta_{0\nu}$ and $\beta\beta_{\sss M}$ decays,
which are the candidates for testing the physics beyond the SM.

In the present work we want
to complete the question of the competition between the standard and exotic
$\beta\beta$-decays \cite{Bar98a}, by scrutinizing the effects of the
nuclear matrix elements, which have been intensively studied in connection
with deviations of the simple-beta spectra from the allowed shape
\cite{Ema61,Sch66,Boh69,Beh82}. (In particular, it should be reminded that
the detection of the
weak-magnetism (WM) term in the spectrum of $\beta^{\pm}$-decays of $^{12}N$
and $^{12}B$ to $^{12}C$ has provided a rather striking test of the CVC
theory.)
It is customary to denominate these HOT as second-forbidden corrections (SFC),
although the most significant among them comes from the WM term,
which obeys the same selection rules as the GT operator.
One also should keep in mind that the WM plays a very important role in both
the $\beta\beta_{0\nu}$ and the $\beta\beta_{\sss M}$ decays, through the
so called {\em recoil term} \cite{Doi85,Bar96,Bar98b}.

Proceeding in the same way as in our previous works \cite{Bar95,Bar98a,Bar98b},
we express the $\beta\beta_{2\nu}$ decay rate as:
\be
d\Gamma_{2\nu}=2\pi \intsum |R_{2\nu}|^2 \delta(\epsilon_1+\epsilon_2
+\omega _1+\omega _2-Q)\prod_{k=1}^2d{\bf p}_k d{\bf q}_k,
\label{2}\ee
where the symbol ${\intsums}$ represents both the summation on lepton spins, 
and the integration on neutrino momenta and electron directions.
The transition amplitude reads \cite{Bar95,Bar98a,Bar98b}:
\br
R_{2\nu}&=&\frac{1}{2(2\pi)^6}\sum_{\ss N}[1-P(e_1e_2)][1-P(\nu_1\nu_2)]
\frac{\bra{0_{\sss F}^+}H_{\sss W}(e_2\nu_2)
\ket{{\ss N}}\bra{{\ss N}}H_{\sss W}(e_1\nu_1)
\ket{0_{\sss I}^+}}{E_{\sss N}-E_{\sss I}+\epsilon_1+\omega _1},
\label{3}\er
where the operator  $P$ exchanges the lepton quantum numbers
$e_i\equiv (\epsilon_i,{\bf p}_i,s_{e_i})$ and
$\nu_i\equiv (\omega _i,{\bf q}_i,s_{\nu_i})$. All other notations have
the usual meaning \cite{Bar98a}.

In the non-relativistic weak Hamiltonian we only retain the terms that
are necessary for the evaluation of the SFC, and obtain
\be
H_{\sss W}(e_i\nu_i)=\frac{G g_{\sss A}}{2}
\left[\mbs+\left(3\xi+\e_i-q_i\right)\Xb
+q_i\left(3\xi+\e_i\right)\Yb +\left(3\xi+\e_i+q_i\right)\Zb \right]
\cdot\Lb(e_i\nu_i),
\label{4}\ee
Here  $G=(2.996\pm 0.002){\times} 10^{-12}$ is the Fermi coupling constant
(in natural units), $\xi=\alpha Z/2R=1.18 ZA^{-1/3}$ is the Coulomb factor,
and
\br
\Xb&=&\frac{g_{\sss V}}{3Mg_{\sss A}}\left(
f_{\sss W}\mbs+\rb\x\pb\right),
\nn\\
\Yb&=&\frac{r^2}{27}\left[\mbs+2\sqrt{8\pi}
(\mbs\otimes Y_2(\hat{\rb}))_1\right],
\label{5}\\
\nn\\
\Zb&=&\frac{1}{6M}[\mbs+2i(\mbs\cdot\pb)\rb],
\nn\er
are the nuclear operators, with $g_{\sss V}=1$ and $f_{\sss W}=4.7$.
The leptonic part is:
\be
\Lb(e_1\nu_i )=sg(s_{\nu_i})
\sqrt{\frac{\epsilon_i+1}{2\epsilon_i} F_0(\epsilon_i)}
\chi^\dagger(s_{e_i})\left(1-\frac{\mbs\cdot{\bf p}}{\epsilon_i+1}\right)\mbs
(1-\mbs\cdot\hat{\bf q})\chi(-s_{\nu_i}).
\label{6}\ee

Keeping only
the interference terms between $\mbs$ with $\Xb$, $\Yb$ and $\Zb$, and the
linear contributions in the lepton energies, from \rf{3} and \rf{4}, we
obtain:
\footnote{ A factor of $3$ has been omitted in the denominator of eq. (8)
of ref. \cite{Bar98a}. All other formulas are, however, correct.}
\be
R_{2\nu}=\frac{G^2}{12(2\pi)^6}
\left[f_0(\e_1\e_2){\cal M}_{2\n}^{\sss (0)}+\frac{1}{2}
\sum_{i=1}^3 f_i(\e_1\e_2){\cal M}_{2\n}^{\sss (i)}\right]
[1-P(\nu_1\nu_2)]\Lb(e_2\n_2)\cdot\Lb(e_1\n_1),
\label{7}\ee
where
\be
f_0(\e_1\e_2)=1,~~~
f_1(\e_1\e_2)=1+2\frac{\e_1+\e_2}{6\xi-Q},~~~
f_2(\e_1\e_2)=1-\frac{\e_1+ \e_2}{Q},~~~~
f_3(\e_1\e_2)=1,
\label{8}\ee
and
\br
{\cal M}_{2\n}^{\sss (1)}&=&
2g^2_{\sss A}(6\xi-Q)\sum_N\frac{\bra{0^+_{\sss F}}
\mbs\ket{1^+_{\sss N}}\cdot\bra{1^+_{\sss N}}\Xb\ket{0^+_{\sss I}}}
{E_{1^+_N}-E_{0^+_I}+Q/2},\nn\\
{\cal M}_{2\n}^{\sss (2)}&=&
6g^2_{\sss A}\xi Q\sum_N\frac{\bra{0^+_{\sss F}}
\mbs\ket{1^+_{\sss N}}\cdot\bra{1^+_{\sss N}}\Yb\ket{0^+_{\sss I}}}
{E_{1^+_N}-E_{0^+_I}+Q/2},
\label{9}\\
{\cal M}_{2\n}^{\sss (3)}&=&
2g_{\sss A}^2\left(6\xi+Q\right)\sum_N\frac{\bra{0^+_{\sss F}}
\mbs\ket{1^+_{\sss N}}\cdot\bra{1^+_{\sss N}}\Zb\ket{0^+_{\sss I}}}
{E_{1^+_N}-E_{0^+_I}+Q/2}.
\nn\er
It might be interesting to note that, while the FFNU
transitions interfere with the GT operator at the level of the transition
rates, the SFC do it already at the level of the transition amplitude.

At the same order of approximation the differential transition rate reads:
\be
d\Gamma_{2\nu}=\frac{4G^4}{15\pi^5}d\Omega_{2\nu}
{\cal M}_{2\n}^{\sss (0)}
\sum_{i=0}^3 f_i(\e_1\e_2){\cal M}_{2\n}^{\sss (i)},
\label{10}\ee
where 
\be
d\Omega_{2\nu}=\frac{1}{64\pi^2}(Q-\epsilon_1-\epsilon_2)^5
\prod_{k=1}^2p_k\epsilon_kF_0(\epsilon_k)d\epsilon_k.
\label{11}\ee
Finally we derive the expressions for the spectrum shape
\be
{d\Gamma_{2\nu}\over d\epsilon}=\frac{G^4}{240\pi^7}
{\cal M}_{2\n}^{\sss (0)}
\sum_{i=0}^3 {\cal F}_i(\epsilon){\cal M}_{2\n}^{\sss (i)},
\label{12}\ee
where
\be
{\cal F}_i(\epsilon) = (Q - \epsilon)^5 f_i(\epsilon) \int_1^{\epsilon-1}
d\epsilon_1 p_1 \epsilon_1 p_2 \epsilon_2 F_0(\epsilon_1)F_0(\epsilon_2).
\ee
\label{13}
For the half-life we get
\be
T_{2\nu}(0_{\sss I}^+{\rightarrow} 0_{\sss F}^+)
=\ln 2[\Gamma_{2\nu}(0_{\sss I}^+{\rightarrow} 0_{\sss F}^+)]^{-1}
=\left({\cal M}_{2\n}^{\sss (0)}
\sum_{i=0}^3{\cal G}_i{\cal M}_{2\n}^{\sss (i)}
\right)^{-1},
\label{14}\ee
with the kinematical factors
\be
{\cal G}_i=\frac{G^4}{240\pi^7\ln 2}\int_2^Q d\epsilon {\cal F}_i(\epsilon).
\label{15}\ee

As seen from \rf{12} the spectrum shape mainly depends on the factors
${\cal F}_i(\epsilon)$. They are displayed for $^{76}Ge$ in the upper panel
of figure 1, as a function of the energy $\epsilon$.
${\cal F}_0(\epsilon)$ and ${\cal F}_3(\epsilon)$ exhibit the same energy
dependence, while ${\cal F}_1(\epsilon)$ shifts the A spectrum slightly to
the right and ${\cal F}_2(\epsilon)$ slightly to the left.

The spectrum shapes without and with the SFC are compared in the lower
panel of figure 1. It can be observed that they are quite similar.
Analogous behavior was found for other experimentally interesting nuclei,
such as $^{82}Se$, $^{100}Mo$, $^{128}Te$, and $^{130}Te$.

\newpage
From \rf{5},\rf{9} and \rf{10} one sees that the main effect of the WM consists
in renormalizing the GT matrix element \rf{1} as:
\be
{\cal M}_{2\nu}^{(0)} \go {\cal M}_{2\nu}^{(0)}
\left(1+\frac{2g_{\sss V}\xi f_{\sss W}}{g_{\sss A}M}\right),
\label{16}\ee
\ie by a factor of $\sim 1.05$ for medium heavy nuclei, {\it independently
of the nuclear model employed.}

The matrix elements ${\cal M}_{2\nu}^{(i)}$ were evaluated within
the pn-QRPA model, following the procedure adopted in our previous works
\cite{Bar95,Bar98b,Krm94}. We display them in table \ref{tab1}, together
with the corresponding kinematical factors  ${\cal G}_i$.
Besides the WM term,
the velocity dependent matrix elements $\rb\x\pb$ and
$2i(\mbs\cdot\pb)\rb$ are also important, and particularly in
the case of $^{100}Mo$.
\footnote{From the theoretical point of view the nuclear
matrix elements in $^{100}Mo$ are in same sense
peculiar, because of the strong predominance of the
$[0g_{7/2}(n)0g_{9/2}(p);J^{\pi}=1^+]$ configuration.}
The moment ${\cal M}_{2\nu}^{(2)}$ is always relatively small.

\begin{table}[h]
\begin{center}
\caption {Numerical results for the kinematical factors ${\cal G}_i$
(in units of $y^{-1}$) and for the nuclear matrix elements
${\cal M}_{2\n}^{\sss (i)}$ (in natural units), evaluated within the
pn-QRPA formalism with an effective axial charge $g_{\sss A}=1$.}
\label{tab1}
\bigskip
\begin{tabular}{|c|c|c|c|c|c|c|r|}
\hline
\hline
Nucleus&${\cal G}_0={\cal G}_3$&${\cal G}_1$&${\cal G}_2$&$
{\cal M}_{2\n}^{\sss (0)}$&${\cal M}_{2\n}^{\sss (1)}$&$
{\cal M}_{2\n}^{\sss (2)}$&${\cal M}_{2\n}^{\sss (3)}$
\\
\hline
$^{76}Ge$&$\!5.49~10^{-20}\!$&$\!6.24~10^{-20}\!$&$\!2.33~10^{-20}\!
$&$\!0.050\!$&$\!0.0044\!$&$\!0.0002\!$&$\!-0.0017\!$\\
$^{82}Se$&$\!1.83~10^{-18}\!$&$\!2.14~10^{-18}\!$&$\!0.83~10^{-18}\!$
&$\!0.060\!$&$\!0.0041\!$&$\!0.0003\!$&$\!-0.0018\!$\\
$^{100}Mo$&$\!3.97~10^{-18}\!$&$\!4.54~10^{-18}\!$&$\!1.83~10^{-18}\!$
&$\!0.051\!$&$\!0.0141\!$&$\!0.0015\!$&$\!0.0072\!$\\
$^{128}Te$&$\!3.54~10^{-22}\!$&$\!3.79~10^{-22}\!$&$\!1.13~10^{-22}\!$
&$\!0.059\!$&$\!0.0048\!$&$\!0.0003\!$&$\!-0.0016\!$\\
$^{130}Te$&$\!2.00~10^{-18}\!$&$\!2.23~10^{-18}\!$&$\!0.91~10^{-18}\!$
&$\!0.048\!$&$\!0.0039\!$&$\!0.0005\!$&$\!-0.0014\!$\\
\hline \hline \end{tabular} \end{center}
\end{table}

As it is well known, within the QRPA the GT moment ${\cal M}_{2\nu}^{(0)}$
strongly depends on the particle-particle coupling constant in the $S=1,T=0$
channel, denoted by $t$ in ref. \cite{Krm94}.
This dependence is particularly pronounced in the physical region for $t$,
where ${\cal M}_{2\nu}^{(0)}$ goes to zero and the QRPA collapses.
${\cal M}_{2\nu}^{(1)}$ and ${\cal M}_{2\nu}^{(3)}$ also strongly depend
on $t$, but in a slightly different way than ${\cal M}_{2\nu}^{(0)}$.
Thus the decay rates rely on two or three rapidly varying functions of $t$,
which makes the calculated half-lives ($T^{\sss A+SFC}_{2\nu}$) to be even
more sensitive on the value of $t$ than within the A approximation
($T^{\sss A}_{2\nu}$). The numerical results are shown in table \ref{tab2}.
Contrarily to what happens in the case of the FFNU transitions,
the SFC decrease the half-lives. The reduction ranges from $6\%$ in
$^{128}Te$ up to $\sim 32\%$ in $^{100}Mo$.

\begin{table}[h]
\begin{center}
\caption {
Calculated half-lives within the allowed approximation
($T^{\sss A}_{2\nu}$) and with second-forbidden corrections included
($T^{\sss A+SFC}_{2\nu}$) in units of $y$.}
\label{tab2}
\bigskip
\begin{tabular}{|c|c|c|}
\hline
\hline
Nucleus
&$T^{\sss A}_{2\nu}$&$T^{\sss A+SFC}_{2\nu}$\\
\hline
$^{76}Ge$
&$\!7.3~10^{21}\!$&$\!6.8~10^{21}\!$\\
$^{82}Se$
&$\!1.5~10^{20}\!$&$\!1.4~10^{20}\!$\\
$^{100}Mo$
&$\!9.7~10^{19}\!$&$\!6.6~10^{19}\!$\\
$^{128}Te$
&$\!8.1~10^{23}\!$&$\!7.6~10^{23}\!$\\
$^{130}Te$
&$\!2.2~10^{20}\!$&$\!2.0~10^{20}\!$
\\
\hline \hline \end{tabular} \end{center}
\end{table}

This work completes our previous inquiries \cite{Bar95,Bar98a} on the
significance of the higher order corrections in the two-neutrino double
beta decay. It can be concluded that:

1) Both the first-forbidden transitions through
the $J^\pi=0^-,1^-$ virtual states, and the second order corrections
to the Gamow-Teller states, affect the transition rates in a significant way.

2) The theoretical uncertainties within the QRPA, in the evaluation of the
half-lives, are of the same order of magnitude
(or even larger) than the experimental errors.

3) The effect on the energy spectra
of the higher order terms in the weak Hamiltonian is too small
to shadow the possible exotic neutrinoless events in contemporary experiments.

\newpage

{\bf Figure Caption}

\vskip 0.5cm

Fig. 1. Kinematical factors ${\cal F}_i$ (upper panel) and the spectrum shape
$d\Gamma_{2\nu}/d\epsilon$ (lower panel) for $^{76}Ge$. All  quantities
are normalized to the maximum value of the allowed shape.

\end{document}